\documentclass[%
reprint,
superscriptaddress,
 amsmath,amssymb,
 aps,
prb,
]{revtex4-1} 

\usepackage{graphicx}

\usepackage[normalem]{ulem}

\usepackage{subfig}

\usepackage{float} 

\usepackage{dcolumn}
\usepackage{bm}
\usepackage{hyperref}
\usepackage{soul} 
\usepackage{xcolor} 

\usepackage[]{natbib}
\definecolor{forestgreen}{rgb}{0.13, 0.55, 0.13}

\usepackage{import} 
\usepackage{labelgraphs}

\def\Re{{\rm Re\mit}}
\def\Im{{\rm Im\mit}}

\begin{document}
 
 \title{Spaser and Optical Amplification Conditions in Graphene-Coated Active Wires}

\author{Leila Prelat}
\affiliation{%
Grupo de Electromagnetismo Aplicado, Departamento de F\'isica, Universidad de Buenos Aires and IFIBA, Consejo Nacional de Investigaciones Cient\'ificas y T\'ecnicas,  Ciudad Universitaria, Pabell\'on I, Buenos Aires 1428, Argentina.
}%
\author{Mauro Cuevas}
\affiliation{
Consejo Nacional de Investigaciones Cient\'ificas y T\'ecnicas and Facultad de Ingenier\'ia (CONICET)
}%
\affiliation{
Universidad Austral, Facultad de Ingenier\'ia, Pilar Mariano Acosta 1611 B1629WWAâ€“Pilarâ€“Buenos Aires Argentina}
\author{Nicol\'as Passarelli}
\affiliation{%
Instituto de Ciencias de Materiales de Barcelona, Consejo Nacional de Investigaciones Cient\'ificas, 08193 Barcelona, Spain.
}%
\author{Ra\'ul Bustos Mar\'un}
\affiliation{%
Instituto de F\'isica Enrique Gaviola, Consejo Nacional de Investigaciones Cient\'ificas y T\'ecnicas and Facultad de Ciencias Qu\'imicas, Universidad Nacional de C\'ordoba, Ciudad Universitaria, C\'ordoba 5000, Argentina.
}%
\author{Ricardo Depine}
\affiliation{%
Grupo de Electromagnetismo Aplicado, Departamento de F\'isica, Universidad de Buenos Aires and IFIBA, Consejo Nacional de Investigaciones Cient\'ificas y T\'ecnicas, Ciudad Universitaria, Pabell\'on I, Buenos Aires 1428, Argentina.
}%


\date{Compiled \today}

\begin{abstract}
This work analyzes the optical properties  of a localized surface plasmon (LSP) spaser made of a  dielectric active wire coated with a graphene monolayer. Our theoretical results, obtained by using rigorous electromagnetic methods, illustrate the non-radiative transfer between the active medium and the localized surface plasmons of the graphene. In particular, we focus on the lasing conditions and the tunability of the LSP spaser in two cases: when the wire is made of an infrared/THz transparent dielectric material and when it is made of a metal-like material. 
We analyze the results by comparing them with analytical expressions obtained by us using the quasistatic approximation.
We show that the studied systems present a high tunability of the spaser resonances with the geometrical parameters as well as with the chemical potential of the graphene.
\end{abstract}



\maketitle

\section{Introduction}
\label{sec:level1}


Spaser, which stands for surface plasmon amplification by stimulated emission of radiation, is the counterpart of a laser that ideally emits surface plasmons instead of photons. Due to the coupling of localized surface plasmons (LSPs) with the electromagnetic radiation, a spaser can confine light at subwavelength scales, which can be used to provide a controllable source of on-demand high-intensity electromagnetic fields beyond the diffraction limit.
Since its proposal in 2003\cite{BergmanStockman2003} and its first experimental demonstrations in 2009,\cite{noginov2009} a wide variety of geometries and materials/metamaterials for the cavities as well as different compounds for the optical-gain medium have been considered. \cite{Rev1,Rev2,spasergraf1,Rev3,Rev4,Rev5,liu2017,pan2017,pan2018,wang2020,krasnok2020}
Since the first experimental demonstrations in 2009, a wide variety of geometries and materials/metamaterials for the cavities as well as different compounds for the optical-gain medium have been considered \cite{BergmanStockman2003,Rev1,Rev2,spasergraf1,Rev3,Rev4,Rev5}.

Despite the development of many advanced spasers, most of the research in this area 
has been focused on natural plasmonic materials such as metals, which have a number of shortcomings, most notably power losses and a fixed charge density which finds applications in the visible and near–infrared range, but not in the mid–infrarred and terahertz (THz) region. 
However, under adequate circumstances, graphene and other 2D materials also 
have the ability to guide surface plasmons while having interesting advantages, such as significantly lower losses and much better tunability. In 2D materials, the constitutive parameters responsible for sustaining plasmon oscillations can be tuned with different techniques, 
including chemical doping and field effects. Besides, highly doped graphene shows  lower losses and much longer plasmon lifetimes compared with conventional noble metals, and the spectral range where graphene plasmons can be excited --covering from microwave to optical frequencies-- is significantly wider than the range covered by metallic plasmons. This has implications for a wide range of applications, including environmental management, detection of  biological and chemical agents and non-invasive medical diagnostics. 

%
%
Motivated by the convenient plasmonic characteristics of graphene and the continuing advances in the fabrication of infrared active optoelectronic materials \cite{infrared1}, a variety of structures with different geometries has been proposed in the literature. For instance, in \cite{spasergraf0} the authors studied the plasmon amplification properties on a plane spaser formed by two dense monolayers of nanocrystal quantum dots deposited onto both sides of a graphene nanoribon. The structure exploits  the spectral tunability via electostatic gating and the high charge carriers mobility in graphene to amplify the low loss surface plasmons in a wide frequency band and with a relatively low threshold.
Regarding spherical geometries, in Ref. \cite{spasergraf2} a spaser formed by a graphene nano-sphere wrapped with two-level quantum dots was theoreticaly studied. In particular, the authors evaluate the ability of the system to launch surface plasmons on a flat graphene sheet placed close to the sphere. 
In Ref. \cite{spasergraf3} the authors propose a 
graphene--metal 
hybrid plasmonic system surrounded by a quantum dots cascade stack as a nano spaser in the infrared. 
In addition, in ref. \cite{ardakani2020} the authors study, within a full quantum approach in the electrostatic regime, a spaser design consisting of a semiconducting cylindrical wire wrapped by graphene and with an active inner core given by a single quantum wire.
Complete literature reviews of the major developments and latest advances in spaser theory, together with a systematic exposition of some of the key results useful to understand the operation of spasers involving both bulk as well as two dimensional plasmonic materials, can be found in Refs. \cite{spasergraf1} (covering up to the year 2016) and \cite{Rev3}. 



%
%
This paper 
studies the non-radiative transfer of energy from the active medium to LSPs in a not-so-explored configuration where a non-magnetic active medium in the form of a circular wire is coated with a monoatomic layer of graphene that separates the wire from an external, passive dielectric medium. 
We consider two different kind of wire cores: nondispersive dielectric, intrinsically nonplasmonic cores, and Drude dispersive, plasmonic cores. In the first case we assume that the core is made of an infrared and THz transparent material. Thus, in this case the graphene layer introduces LSPs 
which, otherwise were absent in the bare wire. 
In the second case, we assume that the core is made of a metal-like material, such as a semiconductor-based nanocrystal \cite{nanocristal2,nanocristal1}, capable of supporting LSPs 
for wavelengths between 3 $\mu$m and 10 $\mu$m. Thus, an hybridization between graphene plasmons and those already existing in the bare wire occurs in this case. 

Based on the resonant behavior of the spaser structure, the LSP characteristics on a graphene cylindrical spaser can be approached in two different but complementary ways. The first one, named the eigenmode approach, involves the  study of  the solutions to the boundary value problem without external sources \cite{CRD},  while the second one, the scattering approach, involves the  study of the electromagnetic response of the structure when it is excited by an external source \cite{maximo1}. The first approach provides the dispersive characteristics of LSPs supported by the structure, such as dispersion curves and damping rates, and the second approach is related to the electromagnetic response 
via quantification of scattering observables 
when LSPs are externally excited. 

In a first stage, we use the eigenmode approach \cite{CRD}, particularly well-suited for finding the critical gain values $[\text{Im}\,\varepsilon_1 ]_c$ 
of the imaginary part of the permittivity of the active medium where the spaser condition is fulfilled.  In the context of the eigenmode approach, these critical gain values can be rigorously obtained by requiring simply that the imaginary part of a modal frequency be zero \cite{smotrova2011,nosich,passarelli2019}. To do so, we use fully retarded methods in all the examples here presented. Besides, we also invoke the quasistatic approximation, valid in the long  wavelength limit, which allows us to obtain analytical expressions that explain the main results obtained with the rigorous theory. 

In a second stage, we use the scattering approach which provides a complementary understanding of plasmonic losses compensation and lasing conditions in terms of scattering observables such as scattering, extinction and absorption cross sections.  

%
%

This paper is organized as follows. In section 2 we present a brief description of the electromagnetic  theory based on separation of variables to obtain both the dispersion equation and the electromagnetic fields associated to LSPs supported  by the structure, and we develop the quasistatic approximation. In section 3  we present numerical examples for  dielectric and metal-like wires covered with a graphene sheet. 
Finally, concluding remarks are provided in Section 4. 
The system of units used is Gaussian and the $\exp(-i\omega t)$ time--dependence is implicit throughout the paper, 
with $\omega$ the angular frequency, $t$ the time, and $i=\sqrt{-1}$. The symbols {\rm Re \mit} and {\rm Im \mit} are used for denoting the real and imaginary parts of a complex quantity respectively. 


\begin{figure}[htbp]
    \centering
    \includegraphics[width=0.35\textwidth]{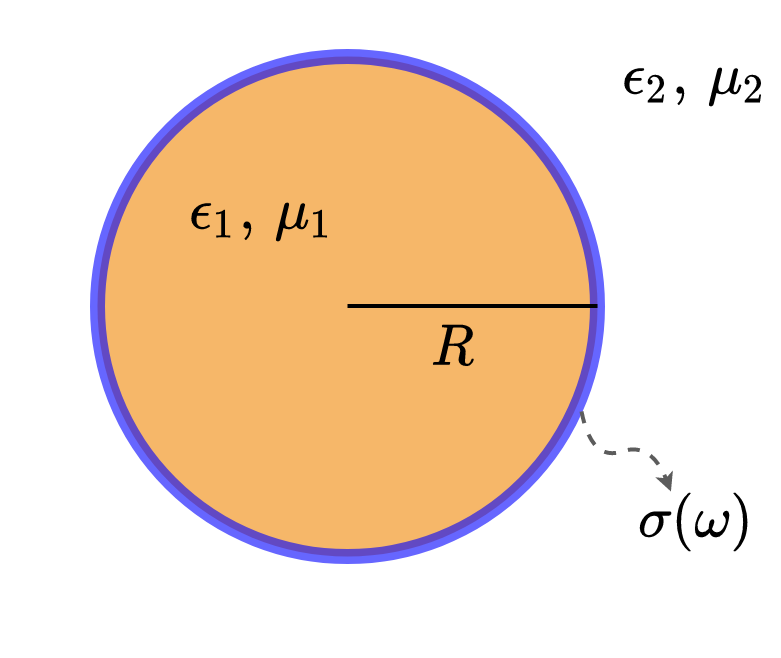}
    \caption{A cylindrical spaser consisting of an active wire coated with a graphene monolayer inmersed in an optically transparent medium.}
    \label{dibujo}
\end{figure}

\section{Theory}
\label{sec:theory}

We consider a graphene--coated cylinder with circular cross--section (radius $R$) centered at $x$=0, $y$=0 (see Fig. \ref{dibujo}) embedded in a transparent medium with real valued electric permittivity $\varepsilon_{2}$ and magnetic permeability $\mu_{2}$. 
The active core is assumed to have a magnetic permeability $\mu_{1}=1 $ and a complex valued electric permittivity 
$\varepsilon_{1}={\rm Re \mit} \,\varepsilon_1+ i\,{\rm Im \mit} \,\varepsilon_1$, with ${\rm Re \mit} \,\varepsilon_1>0$ and ${\rm Im \mit} \,\varepsilon_1<0$. Under these conditions the gain coefficient of the core is $\beta_g=-k_0 \,{\rm Im \mit} \,\varepsilon_1/\sqrt{{\rm Re \mit} \,\varepsilon_1}$ \cite{maier2006}, where $k_0=\omega/c$ is the free space propagation constant. 
The graphene layer is treated as an infinitesimaly thin, local and isotropic two-sided layer with complex valued surface conductivity $\sigma(\omega)$ and we assume that R is large enough, so that the constitutive properties of the graphene coating are the same of those of planar graphene. In this way, we can write 
$\sigma(\omega)=\sigma ^{intra}(\omega) +  \sigma^{inter}(\omega)$, with intraband ($\sigma ^{intra}$) and interband ($\sigma ^{inter}$) transition contributions given by the high frequency expression derived from the Kubo formula (equation (1), Ref. \cite{kubo1}) 
\begin{equation} \label{intra}
\sigma^{intra}(\omega)= \frac{2i e^2 k_B T}{\pi \hbar^2 (\hbar\omega+i\gamma_c)} \mbox{ln}\left[2 \mbox{cosh}(\mu_c/2 k_B T)\right],
\end{equation}  
\begin{eqnarray} \label{inter}
\sigma^{inter}(\omega)= && 
\frac{e^2}{\hbar} \bigg\{   \frac{1}{2}+\frac{1}{\pi}\mbox{arctan}\left[(\hbar\omega-2\mu_c)/2k_BT\right] \nonumber \\
&& -  \frac{i}{2\pi}\mbox{ln}\left[\frac{(\hbar\omega+2\mu_c)^2}{(\hbar\omega-2\mu_c)^2+(2k_BT)^2}\right] \bigg\}, \\ \nonumber 
\end{eqnarray}  
where $\mu_c$ is the chemical potential (controlled with the help of a gate voltage), $T$ the ambient temperature, $\gamma_c$ the carriers scattering rate, $k_B$ the Boltzmann constant, $\hbar$ the reduced Planck constant and $e$ the elementary charge. 
For large doping, $\mu_c\ll k_BT$, the intraband contribution (\ref{intra}) dominates and takes the form predicted by the Drude model, whereas the interband contribution (\ref{inter}) dominates for large frequencies $\hbar \omega > \mu_c$ \cite{kubo1,kubo2}. 

\subsection{Rigorous solution}

In order to derive the modal characteristics of the LSPs in the graphene-coated, circular cross-section wire in terms of wire size, constitutive parameters of substrate and ambient media, and the parameters of the graphene surface conductivity, we use an accurate electrodynamic approach which closely follows the approach of the usual Lorenz–Mie solution for geometries where the radial and angular dependences of the fields can be separated. 
Thus, the electromagnetic field of LSPs in this case can be represented in terms of cylindrical multipole partial waves characterized by discrete frequencies. 
For surface plasmons localized around the cylinder section the problem can be handled in a scalar way since LSPs are only supported in p polarization, that is, when the electric field is parallel to the main section of the wire \cite{maximo1,CRD} and thus can induce in the graphene coating electric currents along the azimuthal direction $\hat \varphi$. 

The magnetic field $\vec H_{n}(\rho,\varphi,t)$ corresponding to the \mbox{$n$-th} LSP mode is written as
\begin{equation} 
\vec H_{n}(\rho,\varphi,t)= F_{n}(\rho,\varphi) \, \exp{(-i\omega_n t)}\,\hat z\,, \label{magnetico}
\end{equation}
where $\omega_n$ is the complex valued modal frequency. Due to carrier relaxation and radiation losses, plasmon oscillations in passive, dissipative media are always damped and thus the relation 
\begin{equation} 
{\rm Im \mit} \;\omega_n < 0 \,, \label{eqn:condomega}
\end{equation}
must be satisfied. Note that this relation holds even when ${\rm Im \mit} \,\varepsilon_1=0$, that is, when the wire interior is a completely transparent dielectric medium. This is due to the
losses given by the emission of radiation which are always
present.
%
%
To find frequencies $\omega_n$ and field distributions $F_n$ associated with the $n$-th LSP mode, $F_n (\rho,\varphi)$ is expanded as series of  cylindrical harmonics in the internal 
and external 
regions. 
\begin{equation} 
F_{n}(\rho,\varphi)=   
\begin{cases} 
   c_{n}\;J_n(k_1\rho)\,\exp{i n\varphi}  \,, \text{\, $\rho < R$,}         \\
   a_{n}\;H_n^{(1)}(k_2\rho) \,\exp{in\varphi}\,, \text{\, $\rho > R$,}
\end{cases}      \label{Fz12p}
\end{equation}
where $a_{n}$ and $c_{n}$ are complex coefficients, $n=1,\, 2,\, \ldots \,\infty$, $k_{j}=\frac{\omega}{c}\sqrt{\varepsilon_{j}\mu_{j}}$ ($j=1, 2$), $c$ is the speed of light in vacuum, and $J_n$ and $H_n^{(1)}$ are the $n$-th Bessel and Hankel functions of the first kind respectively.  
Using the boundary conditions at the graphene layer ($\rho = R$) we get a system of two homogeneous equations for the complex coefficients $a_{n}$ and $c_{n}$, and by requiring that the determinant of this system of equations to be null, we obtain the following dispersion relation for the LSP eigenmodes represented by the cylindrical multipole partial wave \eqref{magnetico} 
\begin{equation}
\mu_2h_n-\mu_1j_n+  i\mu_1\mu_2 \frac{4\pi}{c^2}\sigma  \omega_n R  j_n h_n =0, \label{eq:disp01}
\end{equation}
where $j_n$ and $h_n$ are
\begin{eqnarray}
&& j_n=\frac{J_n'(k_1 R)}{k_1 R \, J_n(k_1 R)} \,\,\,\,\,\,\,\,\,\, h_n=\frac{H_n'^{(1)}(k_2 R)}{k_2 R \, H_n^{(1)}(k_2 R)} 
\end{eqnarray}
The prime denotes the first derivative with respect to the argument of the function.
The eigenfrequency value fixes the relation between amplitudes of the inner and outer region ($c_{n}$  and $a_{n}$ respectively). 
Thus, the spatial part of the electromagnetic field for the \mbox{$n$-th} mode can be written as 
\begin{equation} 
\vec H_{n}=   
\begin{cases} 
J_n(k_1\rho)\,\exp{i n\varphi} \hat z \,, \text{\, $\rho < R$,}         \\
\frac{{\textstyle k_1\,\varepsilon_2}}{{\textstyle k_2\,\varepsilon_1}}\frac{{\textstyle J_n'(k_1R)}}{{\textstyle H_n'^{(1)}(k_2R)}}\;H_n^{(1)}(k_2\rho) \,\exp{in\varphi}\,\hat z\,, \text{\, $\rho > R$,}
\end{cases}      \label{campoph1}
\end{equation}
\begin{equation} 
{\vec E}_n= 
\begin{cases} 
\frac{\textstyle ic k_1}{\textstyle \omega\varepsilon_1}  \Bigl(in \frac{J_n(k_1\rho)}{k_1\rho}\hat{\rho}- J_n'(k_1\rho)\hat{\varphi}\Bigr) \,\exp{i n\varphi} 
\,, \text{\, $\rho < R$,}         \\
\begin{split}
\frac{\textstyle ic k_1}{\textstyle \omega\varepsilon_1} \frac{\textstyle J_n'(k_1R)}{\textstyle H_n'^{(1)}(k_2R)}& \Bigl(in \frac{\textstyle H_n^{(1)}(k_2\rho)}{k_2\rho}\hat{\rho} \,- \\
&\,\,\, H_n'^{(1)}(k_2\rho)\hat{\varphi}\Bigr) \,\exp{i n\varphi} \,, \text{\, $\rho > R$,}
\end{split}
\end{cases}      \label{campope1}
\end{equation}

\subsection{Lasing thresholds}
Optically active medium relies on the stimulated emission of radiation, and that depends, in turn, on the population inversion between an excited and a ground state of the active components of the medium, dyes, quantum dots, rare earth elements, etc. The saturation effects occur when the electromagnetic fields are so intense that a complete population inversion can no longer be sustained by the pumping mechanism, whatever it comes from (radiative or not radiative). The interplay between the electromagnetic field profile given by a certain mode, the dynamics of the population inversion, and even the external field is what ultimately determines the intensity of the electromagnetic fields of optically active systems.
As discussed in several references,\cite{arnold2015,passarelli2016} the consequence of not taking into account saturation effects is that fields go to infinity once optical losses are exactly compensated.
This is not a problem in principle if one is only interested in finding lasing conditions and not the intensity of the electromagnetic fields. Indeed, this property can be used for finding lasing thresholds as divergences of properties such as the scattering coefficient.\cite{passarelli2016,passarelli2019}
Instead, here we use another strategy, which is to find the critical value of the imaginary part of $\varepsilon_1$ for which the modal eigenfrequency $\omega_n$ is real.\cite{smotrova2011,passarelli2019,nosich} However, we are still not taking into account saturation effects, and thus the true intensity of electromagnetic fields at the lasing condition is beyond the scope of this work.

The permittivity of the active medium $\varepsilon_1$ is a frequency-dependent property in general. However, when the response of the system is approximately the same for the frequency interval of interest, the wideband approximation can safely be used. This approximation, which implies taking $\varepsilon_1$ simply as a complex number, is useful for a first exploration of the system as it allows one to calculate the optical response of a system independently of the characteristics of the active medium. As shown for example in Ref. \cite{passarelli2019}, under the appropriate conditions, this approximation provides reliable results for the lasing thresholds.

\subsection{Quasistatic approximation}

When the size of the cylinder is small compared to the wavelength, $R<<\lambda=2\pi c/\omega$, we can use the quasistatic approximation. 
 Using the small argument asymptotic
expansions for Bessel and Hankel functions, the dispersion equation \ref{eq:disp01} is written as \cite{CRD}, 
\begin{equation}
\varepsilon_1+\varepsilon_2=- \frac{4\pi}{\omega}\sigma(\omega) \frac{i}{R}n. \label{eq:disp02}
\end{equation}
Eq. (\ref{eq:disp02}) allows us to obtain analytic expressions for the LSP frequencies of the two cases: the non-dispersive and dispersive interiors. For large doping ($\mu_c >> k_B T$) and relatively low frequencies($\hbar \,\omega<< \mu_c$) the intraband contribution (\ref{intra}) to the surface
conductivity plays the leading role. In this case, complex roots of 
Eq. (\ref{eq:disp02}) admit analytic expressions that can be obtained as follows. 

\subsubsection{Non-dispersive medium}

By substituting the intraband term  (\ref{intra}) into Eq. (\ref{eq:disp02}), an analytical expression for the eigenfrequency for the non dispersive case is obtained:

\begin{equation}
\label{eq:QE non dispersive}
\omega_n = \sqrt{\dfrac{\omega^2_{on}}{\varepsilon_1+\varepsilon_2}-\left(\dfrac{\gamma_c}{2}\right)^2} -i \dfrac{\gamma_c}{2} \approx \dfrac{\omega_{on}}{\sqrt{\varepsilon_1+\varepsilon_2}}-i\dfrac{\gamma_c}{2},
\end{equation}
where $\omega^2_{on} = \dfrac{4e^2\mu_c n }{\hbar^2R}$ is the effective plasma frequency of the graphene coating for the $n$-th mode. By replacing $\varepsilon_1=\Re\,\varepsilon_1+i\Im\,\varepsilon_1$ into Eq. (\ref{eq:QE non dispersive}) and taking into account that $x=\Im\,\varepsilon_1/(\Re\,\varepsilon_1+\varepsilon_2)<<1$, we can expand $\omega_n$ at first order in $x$,  
\begin{equation}
\label{eq:QE non dispersive2}
\omega_n  \approx \dfrac{\omega_{on}}{\sqrt{\Re\,\varepsilon_1+\varepsilon_2}}-i \frac{1}{2} \Big(\gamma_c+\frac{\omega_{on}\,\Im\,\varepsilon_1}{[\Re\,\varepsilon_1+\varepsilon_2]^{3/2}}\Big).
\end{equation}
%
%
%
Note that, within this approximation, the real part of the eigenfrequency does not depend on the imaginary part of $\varepsilon_1$. In addition, from Eq. (\ref{eq:QE non dispersive2}) we see that the critical value of the imaginary part of $\varepsilon_1$ for which the modal eigenfrequency is real is written as,  
\begin{equation}
\label{eq:im epsi1 crit 2 QE non dispersive}
[\text{Im}\,\varepsilon_1 ]_c = -\dfrac{[\Re\,\varepsilon_1+\varepsilon_2]^{3/2}\, \gamma_c}{ \omega_{on}}=-\dfrac{[\Re\,\varepsilon_1+\varepsilon_2]^{3/2}\, \gamma_c \hbar \, R}{\sqrt{4 e^{2}\mu_c \,n } }    
\end{equation}
%










\subsubsection{Dispersive medium}

We consider a mix of nanocrystal and a dye (active medium) for the interior medium. For the metal-like behavior of the nanocrystal we used the Drude model:

\begin{equation}
\label{eq drude lorentz}
\varepsilon_{DL}(\omega) = \varepsilon_\infty - \dfrac{\omega^2_p}{\omega^2 + i\gamma_m \omega}    
\end{equation}
where $\varepsilon_\infty$ is the residual high-frequency response of the material, $\omega_p$ the metallic plasma frequency and $\gamma_m$ the optical loss rate of the Drude material. Therefore, the effective homogenized permittivity of the medium inside the cylinder is:

\begin{align}
\label{eq permeabilidad 1 disp}
\varepsilon_1(\omega) = \varepsilon_\infty - \dfrac{\omega^2_p}{\omega^2 + i\gamma_m \omega} + \underbrace{\varepsilon_{dr} + i\varepsilon_{di}}_{\text{dye}} = \nonumber \\
 \varepsilon'_\infty - \dfrac{\omega^2_p}{\omega^2 + i\gamma_m \omega} + i\varepsilon_{di}
\end{align}
where $\varepsilon'_\infty \equiv \varepsilon_\infty + \varepsilon_{dr}$ and $\varepsilon_{dr} + i\varepsilon_{di}$ represents the contribution of the dye to the effective homogenized permittivity. Note that in this mixed model of the active medium, only the dye is taken in a wideband approximation.

Replacing into eqaution (\ref{eq:disp02}) the expression of $\varepsilon_1$ given by equation (\ref{eq permeabilidad 1 disp}), after expanding in powers of $y=\frac{\varepsilon_{di}}{\varepsilon'_\infty+\varepsilon_2}$, we obtain, 
%
\begin{align}
\label{eqn:QE dispersive2}
\omega_{n}= \sqrt{\dfrac{\omega^2_p+\omega^2_{0n}}{\varepsilon'_\infty+\varepsilon_2}}-\dfrac{i}{2} \nonumber\\ 
\times \Big[ \dfrac{\omega^2_p\gamma_m+\omega^2_{0n}\gamma_c}{(\omega^2_p+\omega^2_{0n})}+  \varepsilon_{di} \dfrac{\sqrt{\omega^2_p+\omega^2_{0n} }}{(\varepsilon'_\infty+\varepsilon_2)^{3/2}} \Big]
\end{align}
%
%
Similar to the non dispersive case, the real part of the modal eigenfrequencies does not depend on  $\varepsilon_{di}$ within the range of validity of the quasistatic approximation approximation. Moreover, by equating to zero the imaginary part in Eq. (\ref{eqn:QE dispersive2}), we obtain the critical value for $\varepsilon_{di}$,
\begin{equation}
\label{eqn: frecuency crit QE dispersive}
[\varepsilon_{di}]_c = - \dfrac{\omega^2_p\gamma_m+\omega^2_{0n}\gamma_c}{(\omega^2_p+\omega^2_{0n})^{3/2}}(\varepsilon'_\infty+\varepsilon_2)^{3/2}
\end{equation}
for which the lasing condition for the $n$-th mode is reached. 

%








\section{Results}

In this section we solve numerically the fully retarded dispersion relation Eq. (\ref{eq:disp01}) to obtain both the complex eigenfrequencies and the lasing thresholds for the first four multipolar modes. The calculation of the lasing thresholds requires to find the critical values for which the imaginary part of the eigenfrequency is zero, that is, the values which exactly compensate the plasmon losses. 
To do so, we minimize the modulus of equation \eqref{eq:disp01} with respect to two  variables, namely, the real part of the modal frequency and the optical gain. We use the Nelder-Mead optimization algorithm, taking the values provided by the 
quasistatic analytical expressions as initial guesses. We have considered that the wire is immersed in vacuum,  $\varepsilon_2=\mu_2=1$, and that the graphene parameters are $\gamma_c=0.1$meV and T=$300$K in all the calculations. 


\subsection{Non dispersive medium}

\begin{figure}[H]
\centering
\includegraphics[width=0.48\textwidth]{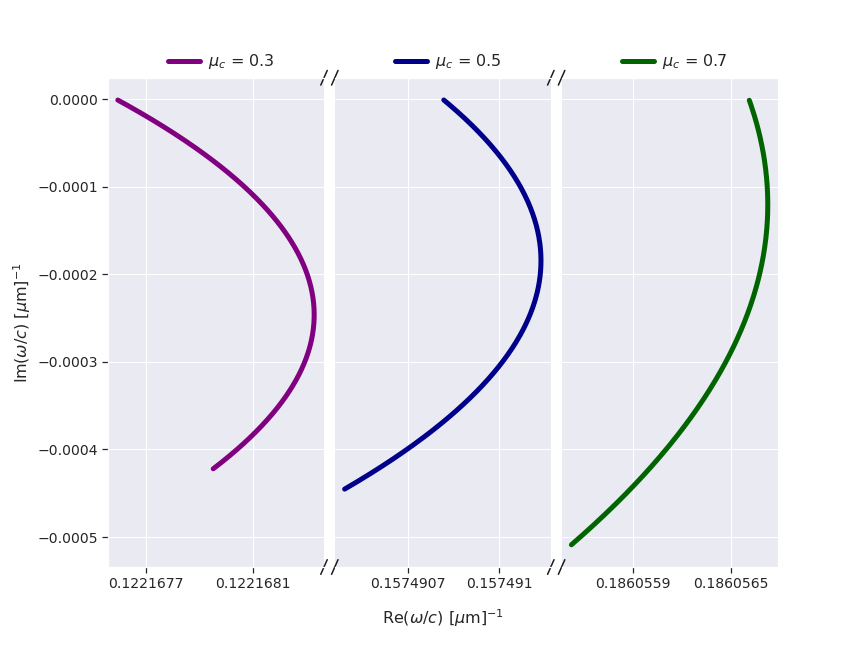}
\caption{Parametric curves of the complex poles (dipolar mode) for a cylinder with  \oneNonDisp $\,\,$ and three values of $\mu_c$.} \label{fig complex poles non disp}
\end{figure}

We consider a dielectric wire of radius $R=0.5\mu$m and $\Re\,\varepsilon_1=4.9$.
In Fig. \ref{fig complex poles non disp}, we plotted the numerical solutions of the dispersion relation, Eq. \eqref{eq:disp01}, for the dipolar mode and for three different values of chemical potential $\mu_c = [0.3,0.6,0.9]$ eV. The solutions are displayed parametrically in the complex plane $\Re\,\omega/c$ - $\Im\,\omega/c$, with the imaginary part of $\varepsilon_1$ as parameter. The lowest part of the curves in Fig. \ref{fig complex poles non disp} corresponds to a passive medium ($\Im\,\varepsilon_1= 0$). These curves approach the real axis when the value of $\Im\,\varepsilon_1$ approaches the  critical value $[\Im\,\varepsilon_{1}]_c$ for which the lasing condition for the dipolar order is reached. We have verified that the curves cross the real axis when $|\Im\,\varepsilon_1|> |\Im\,\varepsilon_{1}|_c$. 
%
%
%
%
The curves in Fig. \ref{fig complex poles non disp} clearly show that, 
for the values of $\Im\,\varepsilon_1$ considered, the real part of the modal frequency remains almost constant. 
This behavior can be understood from the quasistatic expression (\ref{eq:QE non dispersive2}), where the real part of the eigenfrequency does not depend on the   imaginary part of $\varepsilon_1$. 
In addition, we observe that the curves move towards higher $\Re\,\omega/c$ values when the value of the chemical potential is increased, a fact that blueshifts the lasing frequency. This behavior is also predicted by the quasistatic expression (\ref{eq:QE non dispersive2}), which shows that $\Re\,\;\omega_n$ behaves like $\sqrt{\mu_c}$.



\begin{figure}[H]
\centering
\includegraphics[width=0.48\textwidth]{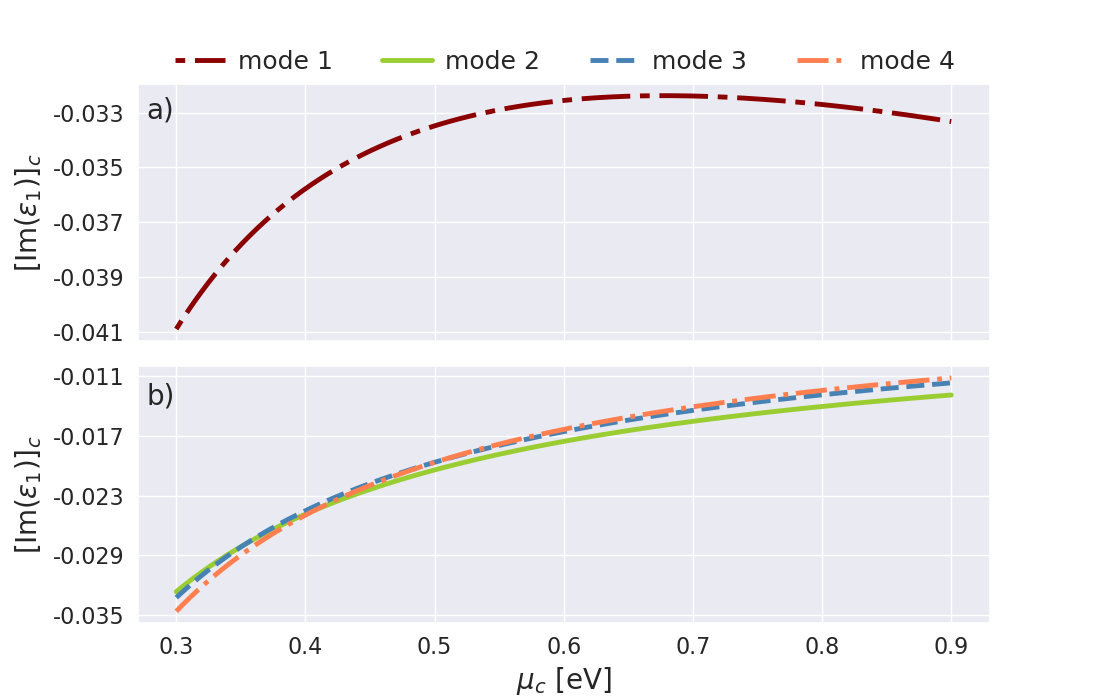}
\caption{Critical values of ${\rm Im \mit} \,\varepsilon_1$ as a function of $\mu_c$ for 
a cylinder with \twoNonDisp. a) The dipolar. b) the quadrupolar, hexapolar and octupolar modes.}
\label{fig critical values non disp}
\end{figure}

Fig. \ref{fig critical values non disp} shows the  critical values  $[\Im\,\varepsilon_{1}]_c$ for which the lasing condition is reached as a function of the chemical potential of graphene for the dipolar mode (Fig. \ref{fig critical values non disp}a) and for quadrupolar, hexapolar and octupolar modes  (Fig. \ref{fig critical values non disp}b).  
From these curves 
we see that the critical gain coefficient $\beta_g=-k_0\, \Im\,\varepsilon_1/\Re\,\varepsilon_1$ for which the dipolar mode reaches the lasing threshold is greater than that corresponding to the other modes. 
This is due to the fact that  radiation losses for the dipolar mode are higher than for  other modes (see \cite{CRD} Table 1), a behavior not predicted by the quasistatic approximation (\ref{eq:QE non dispersive}).
%
%
On the other hand, Eq. (\ref{eq:im epsi1 crit 2 QE non dispersive}) anticipates that in the quasistatic approximation, the critical gain parameter behaves like $\mu^{-1/2}$.  We observe that while the numerical curves for the higher (quadrupolar, hexapolar and octupolar) modes in Fig. \ref{fig critical values non disp}b exhibit this behavior, the numerical curve for the dipolar mode in Fig. \ref{fig critical values non disp}a does not.
This is in accordance with the fact that the higher the multipole modal frequency $\omega_n$, the better the 
quasistatic approximation, 
since the effective wavelength of higher multipoles becomes shorter and the LSP modes perceive the circular graphene sheet as increasingly flat (see Ref. \cite{CRD}, equations (11)-(13)).

\begin{figure}[t]
    \centering
    \captionsetup[subfigure]{justification=centering}
    \subfloat[Scattering]{\includegraphics[width=0.22\textwidth]{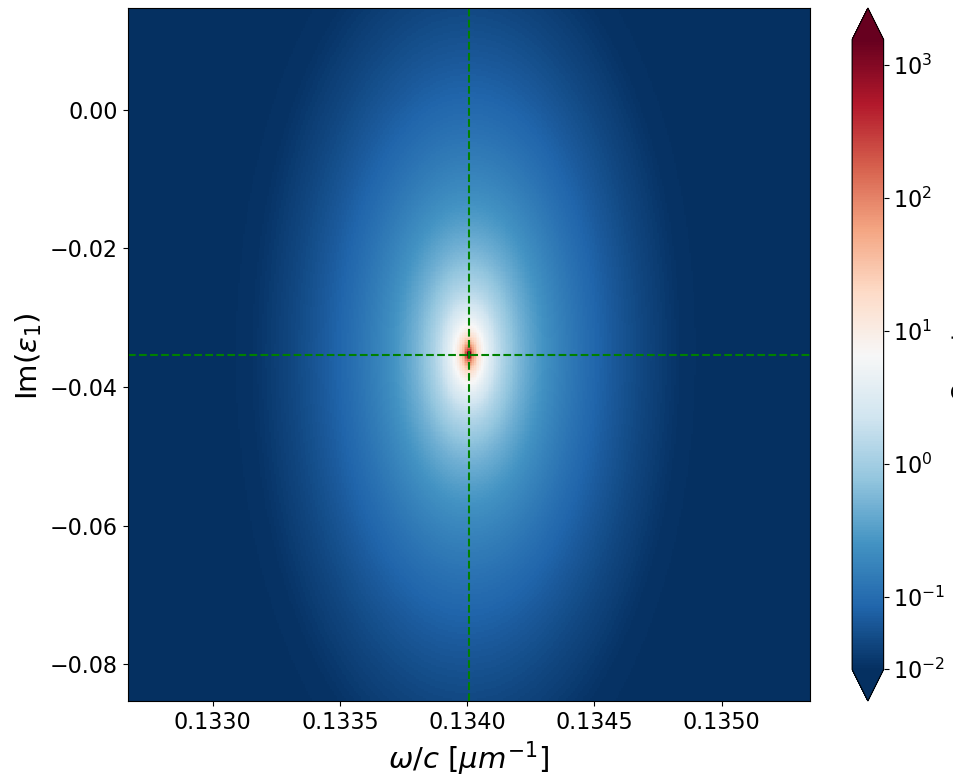}\label{fig Qscat non disp}}
    \captionsetup[subfigure]{justification=centering}
    \subfloat[Absorption]{\includegraphics[width=0.22\textwidth]{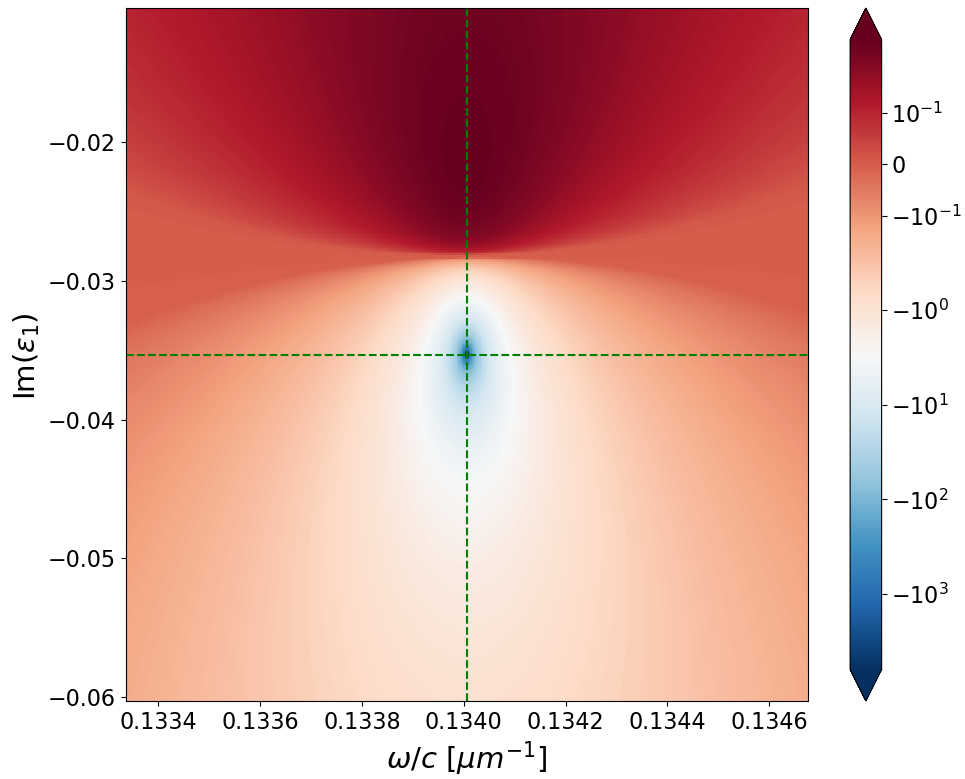}\label{fig Qabs non disp}}
    \captionsetup[subfigure]{justification=centering}
    \subfloat[Extinction]{\includegraphics[width=0.22\textwidth]{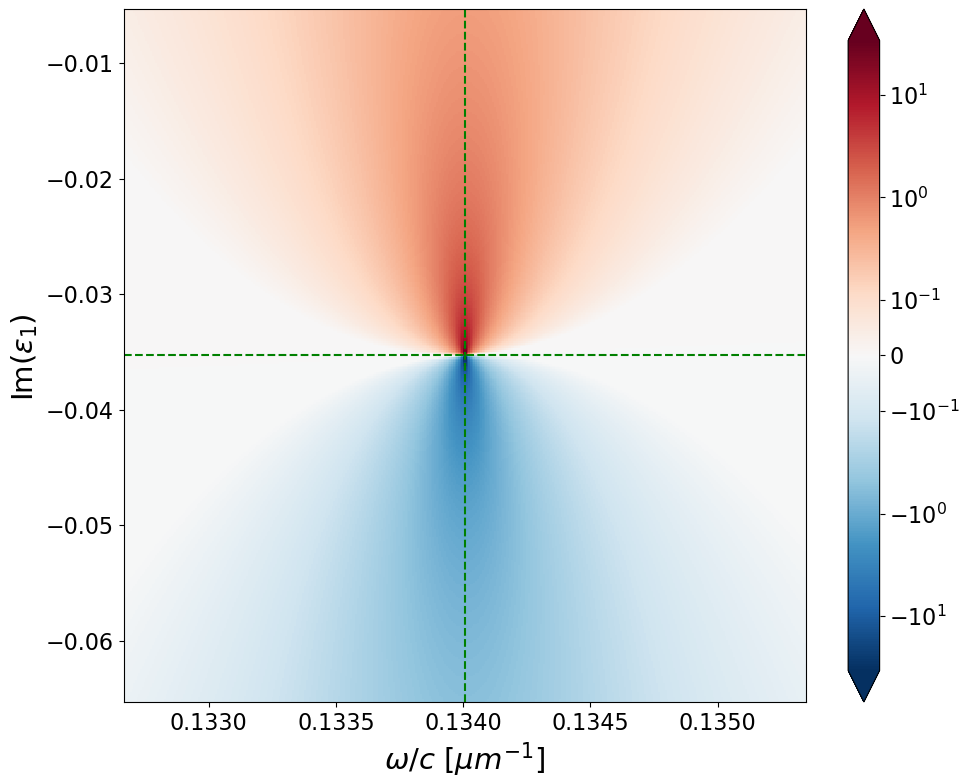}\label{fig Qext non disp}}
    \caption{Cross-section for a cylinder with \threeNonDisp. In green dotted lines: the critical values found before (Fig. \ref{fig critical values non disp})}
    \label{fig cross sections nondisp}
\end{figure}

In order to discuss the previous results in terms of scattering observables, 
in Fig.s \ref{fig Qscat non disp}, \ref{fig Qabs non disp} and  \ref{fig Qext non disp} 
we plot color maps in the $\omega/c$ - $\Im\,\varepsilon_1$ plane of the scattering, extinction and absorption cross sections, respectively. 
Fig. \ref{fig cross sections nondisp}a illustrates the enhancement of the scattering efficiency for frequencies  and gain parameters near the values $\omega_c/c=0.134$ and $[\Im\,\varepsilon_{1}]_c = -0.0353072$ for which the lasing condition for the dipolar mode is reached. Near this critical condition, the scattering cross section diverges and the width at half maximum of the resonance tends to zero, 
in agreement with the fact  that the eigenfrequency tends to be real. 


Fig. \ref{fig Qext non disp} shows two regions separated by a  green  curve  for  which  the  extinction  cross  section  is  equal to zero. Above this curve the extinction cross section is possitive (red  region),  indicating  that  plasmonic  losses  (ohmic  more radiative losses) are not compensated, while below this curve the extinction cross section is negative, indicating  that  plasmonic losses are fully compensated. 
The critical point $\omega_c/c$ and $[\Im\,\varepsilon_{1}]_c$ falls on the full loss compensation curve, since the lasing condition implies the full loss compensation condition \cite{Rev1}. 
From Fig. \ref{fig Qabs non disp} we see that the region for which the absorption cross section is positive falls above the full loss compensation curve, indicating that ohmic losses are compensated for an optical gain coefficient value which is lower than that corresponding to the critical value. This is an indication of deviations from the quasistatic approximation, which does not present radiation losses. 



\begin{figure}[ht!]
    \centering
    \captionsetup[subfigure]{justification=centering}
    \subfloat[mode 1, $\Im\,\varepsilon_1 \neq 0$]
    {\includegraphics[width=0.2\textwidth]{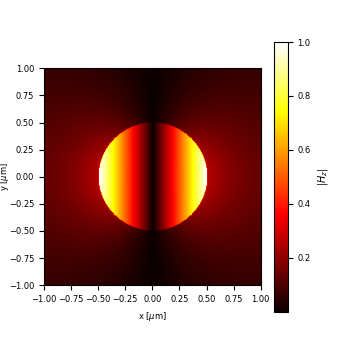}}
    \hspace{1mm}
    \captionsetup[subfigure]{justification=centering}
    \subfloat[mode 1, $\Im\,\varepsilon_1 = 0$]
    {\includegraphics[width=0.2\textwidth]{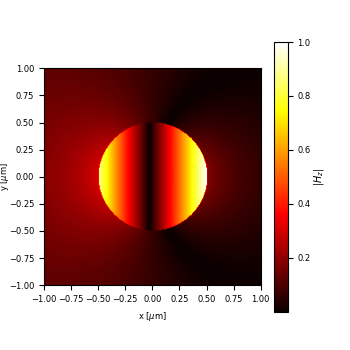}} 
    \newline
    \captionsetup[subfigure]{justification=centering}
    \subfloat[mode 2, $\Im\,\varepsilon_1 \neq 0$]
    {\includegraphics[width=0.2\textwidth]{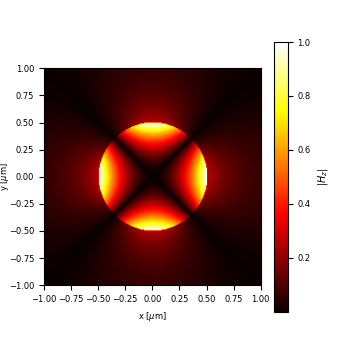}}
    \hspace{1mm}
    \captionsetup[subfigure]{justification=centering}
    \subfloat[mode 2, $\Im\,\varepsilon_1 = 0$]
    {\includegraphics[width=0.2\textwidth]{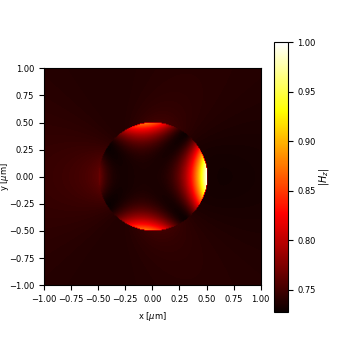}} 
    \newline
    \captionsetup[subfigure]{justification=centering}
    \subfloat[mode 3, $\Im\,\varepsilon_1 \neq 0$]
    {\includegraphics[width=0.2\textwidth]{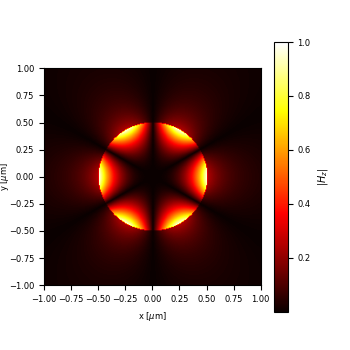}}
    \hspace{1mm}
    \captionsetup[subfigure]{justification=centering}
    \subfloat[mode 3, $\Im\,\varepsilon_1 = 0$]
    {\includegraphics[width=0.2\textwidth]{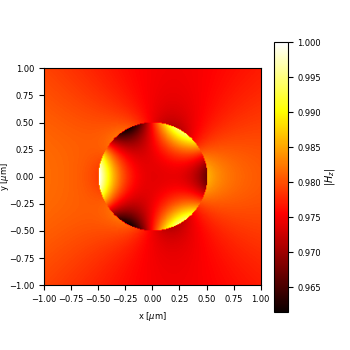}} 
    \newline
    \captionsetup[subfigure]{justification=centering}
    \subfloat[mode 4, $\Im\,\varepsilon_1 \neq 0$]
    {\includegraphics[width=0.2\textwidth]{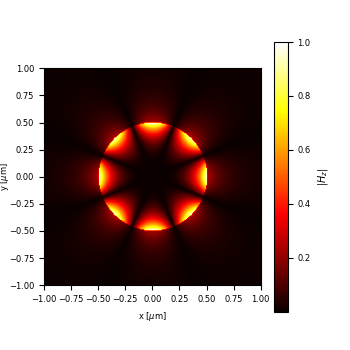}}
    \hspace{1mm}
    \captionsetup[subfigure]{justification=centering}
    \subfloat[mode 4, $\Im\,\varepsilon_1 = 0$]
    {\includegraphics[width=0.2\textwidth]{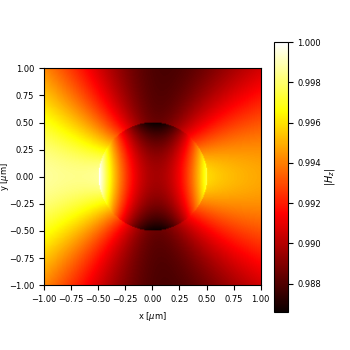}} 
    \caption{$|Hz|$ field with \fourNonDisp}
    \label{fig fields nondisp}
\end{figure}

To gain insight about the active medium inclusion on the electromagnetic field scattered by the wire, in Fig. \ref{fig fields nondisp} we plotted the spatial distribution of the magnetic field $H_z$, at the resonance frequency for the first four modes (dipolar, cuadrupolar, hexapolar and octupolar),  near the wire with an active medium ($\Im\,\varepsilon_1 \neq 0$, left column) and without an active medium ($\Im\,\varepsilon_1 = 0$, right column).  The direction of the plane wave incidence is 
from  left to right.
Each modal field has been  normalized with respect to its own maximum field. 


In Fig \ref{fig fields nondisp}, for the active medium we used the critical values $[\Im\,\varepsilon_{1}]_c = -0.0353072, -0.0271857, -0.0282580, -0.0298547$ for the imaginary part of the permittivity of the active medium and $\omega_c/c =  [0.134005, 0.189642, 0.232135, 0.267846] $  $\mu m^{-1}$ for the dipolar, quadrupolar, hexapolar and octupolar modes.


By comparing the left and right columns of Fig.\ref{fig fields nondisp}, obtained by evaluating the field distribution at each resonant frequency, we observe important differences when the active medium takes a gain value close to the critical one, $[\Im\,\varepsilon_{1}]_c$.
These differences come from the interference with the incident plane wave. Basically, for gains far from the critical value, there is always interference between the fields produced by a given mode and the external source. However, close to the lasing threshold, the fields produced by the eigenmodes are so strong that those coming from the external illumination become negligible.




\subsection{Dispersive medium}

We present rigorous numerical results obtained when the cylinder core is a 
metal-like material, with dielectric permittivity $\varepsilon_1(\omega)$ described by equation (\ref{eq permeabilidad 1 disp}). 
We use $\varepsilon_\infty=3.9$, plasma frequency $\hbar \omega_p=0.6$eV and collision frequency $\hbar \gamma_{m}=0.01$eV. 
%
\begin{figure}[h] 
\centering
\includegraphics[width=0.49\textwidth]{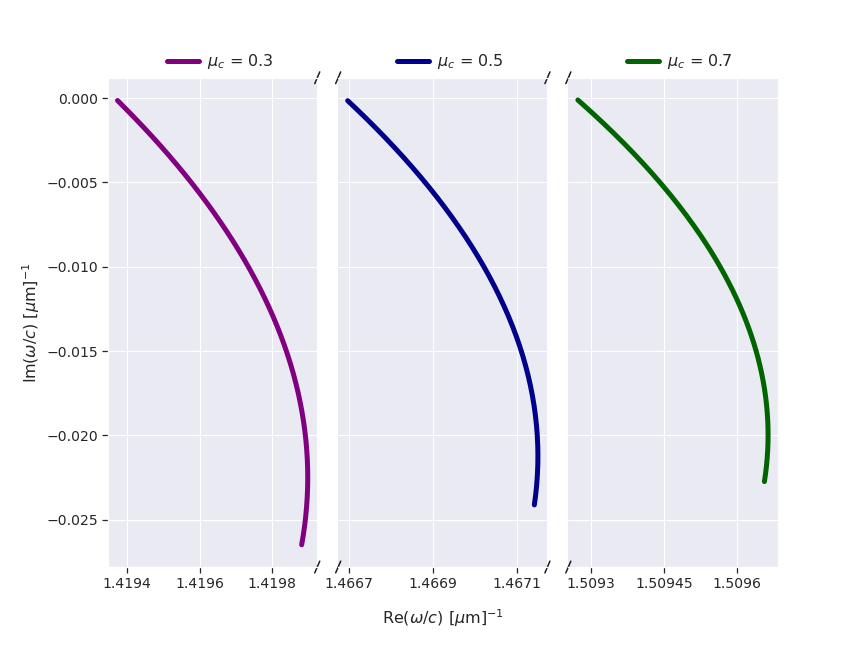}
\caption{Complex poles (dipolar mode) for a cylinder with \oneDisp}
\label{fig complex poles disp}
\end{figure}

In Fig. \ref{fig complex poles disp} we show parametric curves in the 
complex frequency plane of the dipolar eigenfrequency calculated by solving the dispersion relation \eqref{eq:disp01}, as a parametric function of $\varepsilon_{di}$ for three different values of chemical potential $\mu_c = [0.3,0.6,0.9]$ eV. 
%
%
As in the non-dispersive case, a very small variation in the real part of the eigenfrequency is observed, 
which is consistent with the fact that the expression for the real part of the eigenfrequency predicted by the quasistatic approximation (\ref{eqn:QE dispersive2}) is independent of $\varepsilon_{di}$. 
Moreover, we observe that the frequency region for the eigenfrequency trajectory is blue shifted when $\mu_{c}$ is increased from 0.3eV to 0.7eV. This is consistent with the fact that the hybridization formula (\ref{eqn:QE dispersive2}), obtained from the quasistatic  approximation, predicts a resonance frequency (real part of the eigenfrequency) which is proportional to $\sqrt{\omega_p^2+k\mu_c}$, $k=4 e^2/(\hbar^2 R)$.



\begin{figure}[ht!]
\centering
\captionsetup[subfigure]{justification=centering}
\subfloat[R = 0.5 $\mu$m]{\includegraphics[width=0.4\textwidth]{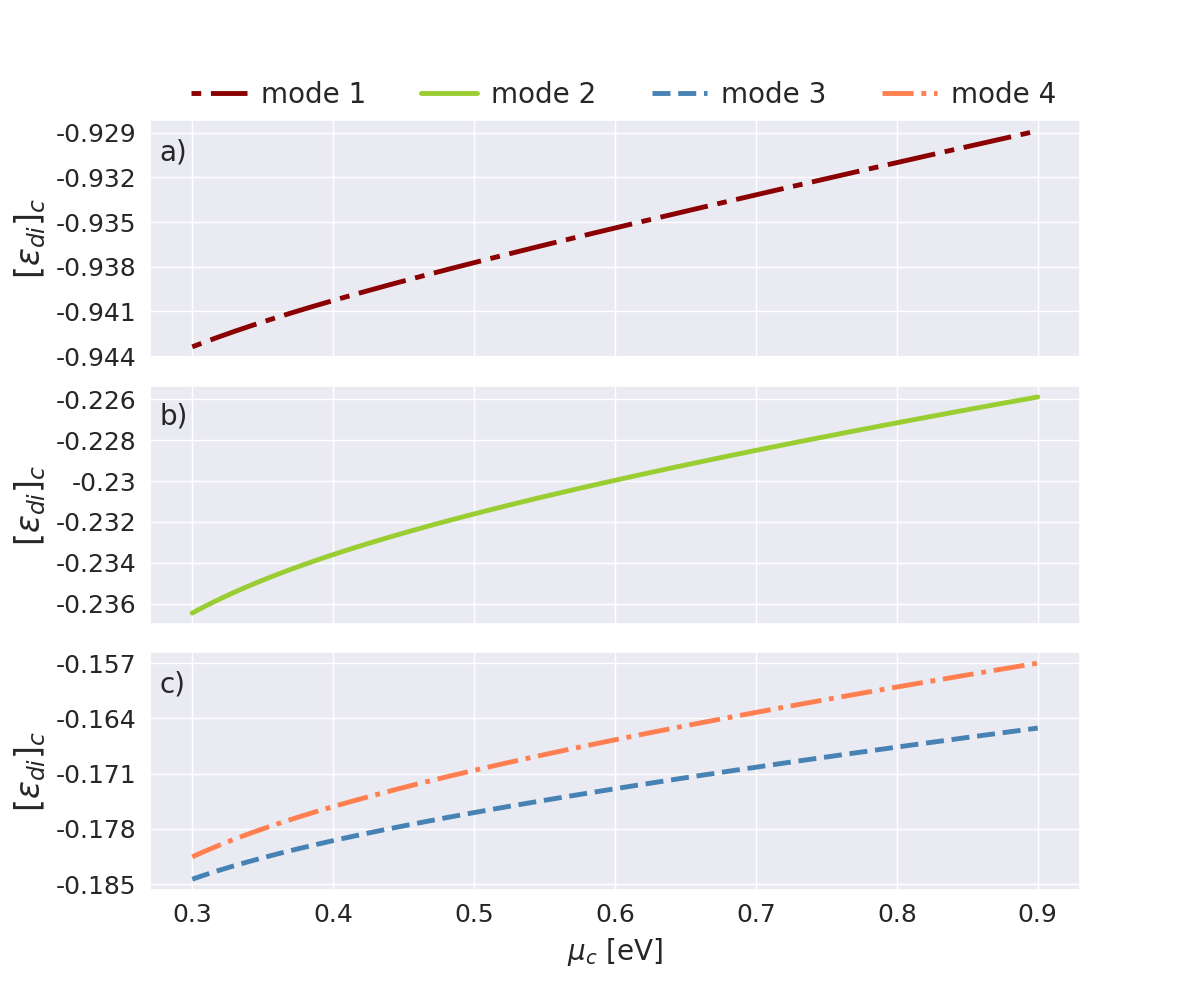}}
\newline
\captionsetup[subfigure]{justification=centering}
\subfloat[R = 0.05 $\mu$m]{\includegraphics[width=0.4\textwidth]{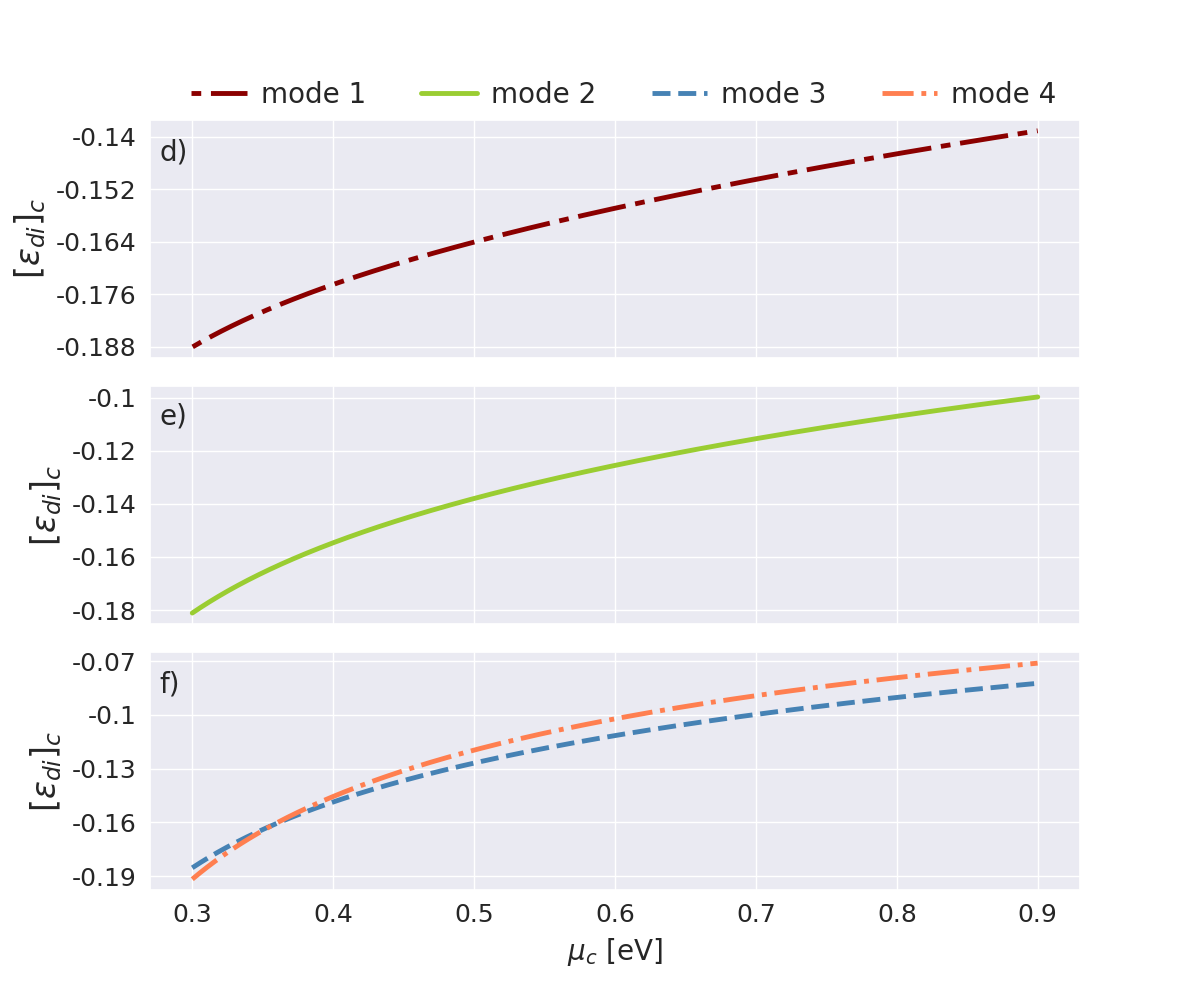}}
\caption{Critical values of $\varepsilon_{di}$ as a function of $\mu$ for for a cylinder with \twoDisp. 
}
\label{fig critical values disp}
\end{figure}

Fig. \ref{fig critical values disp} shows the critical values $[\varepsilon_{di}]_c $ for the first four modes as a function of chemical potential $\mu_c$ for two wires sizes: $R=0.5\mu$m (Fig. \ref{fig critical values disp}a) and $R=0.05\mu$m (Fig. \ref{fig critical values disp}b). 
Unlike the non dispersive case, where the active medium only has to compensate for plasmon losses in the graphene monolayer, in metal-like dispersive cores the active medium has to compensate for plasmon losses both in the graphene layer and in the nanocrystal. Thus, the modulus of the critical values $[\varepsilon_{di}]_c $ for all the graphene-nanocrystal hybridized plasmon modes shown in Fig. \ref{fig critical values disp} are bigger 
than those for the graphene plasmons obtained in the non-dispersive case (Fig. \ref{fig critical values non disp}). 
%

Regarding wire size, Fig. \ref{fig critical values disp} clearly shows that the critical values corresponding to $R=0.05\mu$m (Fig. \ref{fig critical values disp}b) are notably lower than those corresponding to $R=0.5\mu$m (Fig. \ref{fig critical values disp}a). This reduction of the value of the critical gain with particle size 
is anticipated by the quasistatic expression (\ref{eqn: frecuency crit QE dispersive}), which shows that  $|[\varepsilon_{di}]_c|$ is an increasing function of $R$ (note that the dependence on $R$ is included in $\omega^2_{0n}$). It is interesting to note that this behavior, which is absent in metallic cylinders without a graphene cover,  
results from the graphene-nanocristal plasmon hybridization and is highlighted when the effective plasma frequency $\omega_{on}$ is comparable with the nanocrystal plasma frequency $\omega_p$ (as is the case for the constitutive parameters chosen in this  example). 

%
%
%
%





\begin{figure}[H]
    \centering
    \captionsetup[subfigure]{justification=centering}
    \subfloat[Scattering]{\includegraphics[width=0.22\textwidth]{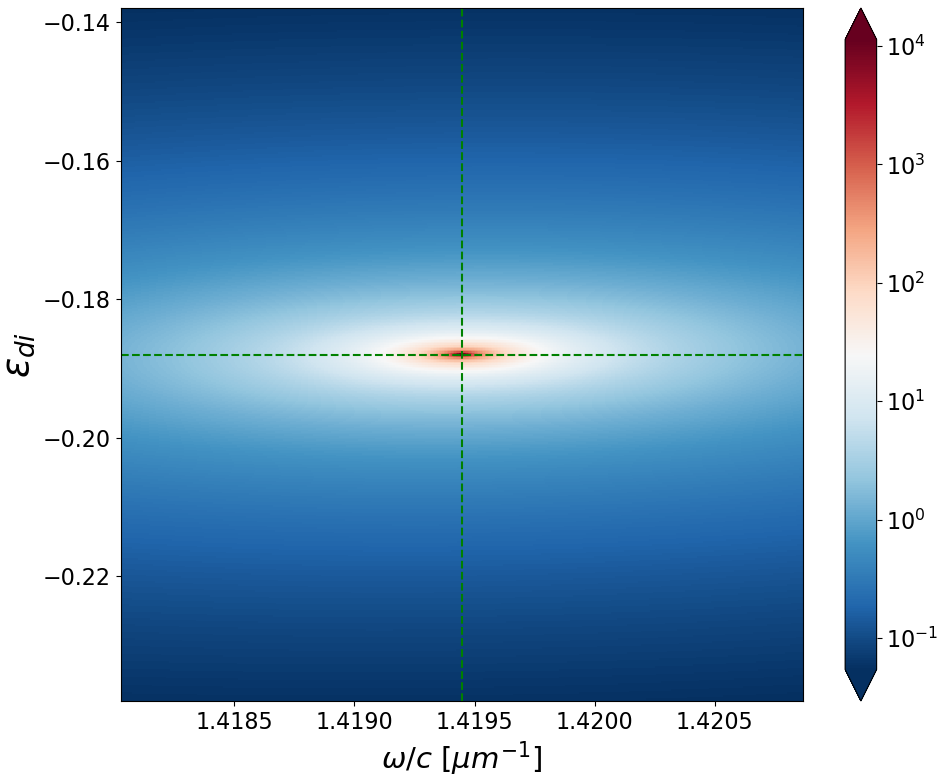}\label{fig Qscat disp}}
    \captionsetup[subfigure]{justification=centering}
    \subfloat[Absorption]{\includegraphics[width=0.22\textwidth]{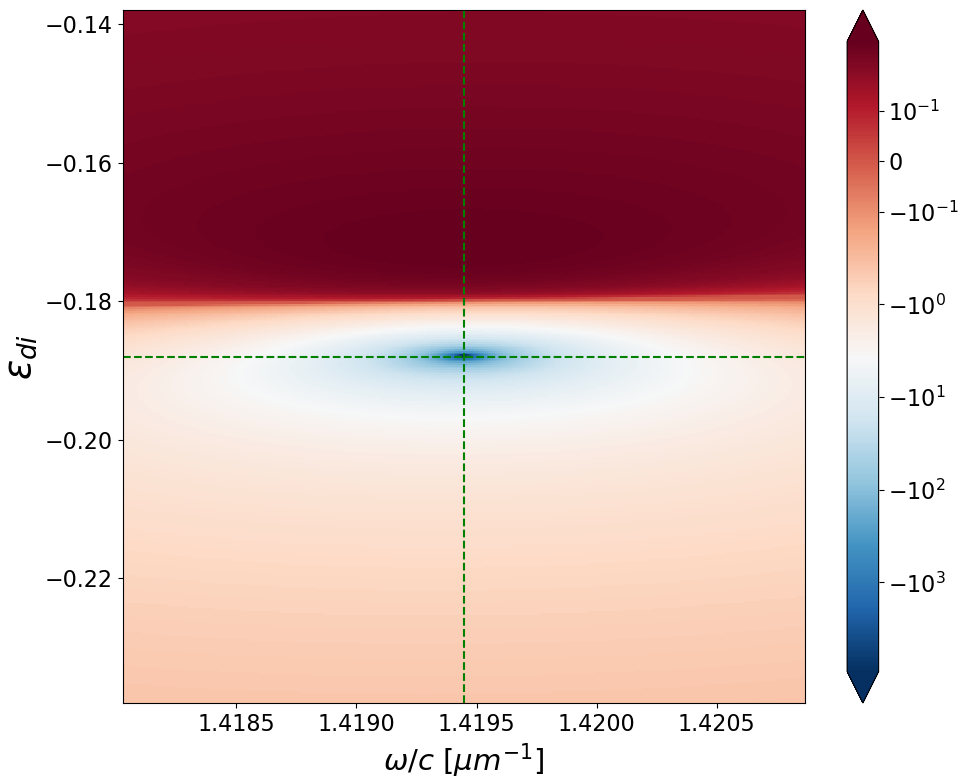}\label{fig Qabs disp}}
    \captionsetup[subfigure]{justification=centering}
    \subfloat[Extinction]{\includegraphics[width=0.22\textwidth]{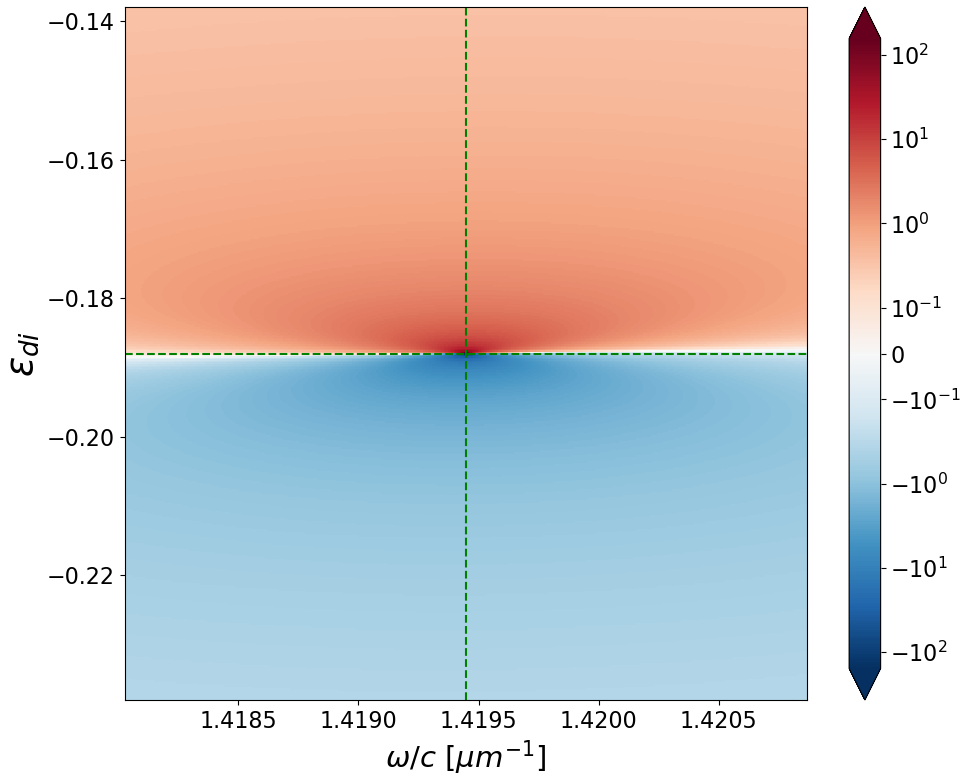}\label{fig Qext disp}}
    \caption{Cross-section for a cylinder with \threeDisp. In green dotted lines: the critical values found before.}
    \label{fig cross sections disp}
\end{figure}

In Fig. \ref{fig cross sections disp} we plot color maps, in the 
$\omega/c$ - $\Im\,\varepsilon_1$ plane, of the scattering (Fig. \ref{fig Qscat disp}), 
extinction (Fig. \ref{fig Qabs disp}) and absorption (Fig. \ref{fig Qext disp}) cross sections. 
We observe that the behavior around the critical point ($\omega_c/c$, $[\varepsilon_{di}]_c$)  is similar to that observed in the non-dispersive case (Fig. \ref{fig cross sections nondisp}), where the critical point falls on the full loss  compensation curve that separates the passive region (red region in Fig. \ref{fig Qext disp}) from the active region (blue region in Fig. \ref{fig Qext disp}).  

\begin{figure}[ht!]
    \centering
    \captionsetup[subfigure]{justification=centering}
    \subfloat[mode 1, $\varepsilon_{di} \neq 0$]
    {\includegraphics[width=0.2\textwidth]{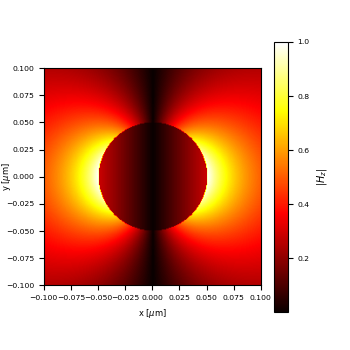}}
    \hspace{1mm}
    \captionsetup[subfigure]{justification=centering}
    \subfloat[mode 1, $\varepsilon_{di} = 0$]
    {\includegraphics[width=0.2\textwidth]{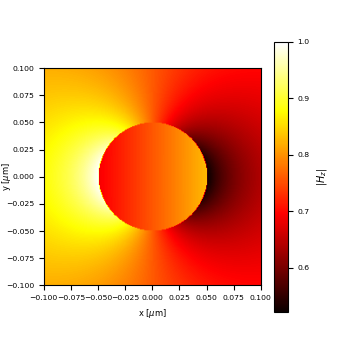}} 
    \newline
    \captionsetup[subfigure]{justification=centering}
    \subfloat[mode 2, $\varepsilon_{di} \neq 0$]
    {\includegraphics[width=0.2\textwidth]{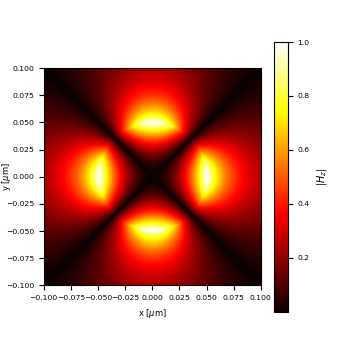}}
    \hspace{1mm}
    \captionsetup[subfigure]{justification=centering}
    \subfloat[mode 2, $\varepsilon_{di} = 0$]
    {\includegraphics[width=0.2\textwidth]{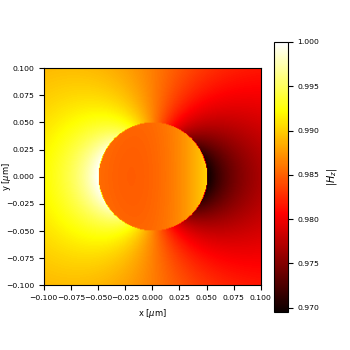}} 
    \newline
    \captionsetup[subfigure]{justification=centering}
    \subfloat[mode 3, $\varepsilon_{di} \neq 0$]
    {\includegraphics[width=0.2\textwidth]{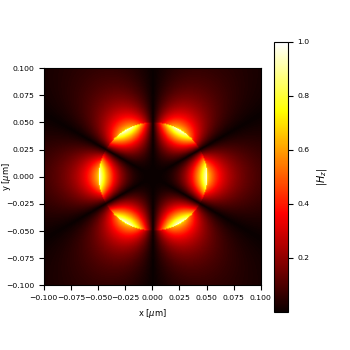}}
    \hspace{1mm}
    \captionsetup[subfigure]{justification=centering}
    \subfloat[mode 3, $\varepsilon_{di} = 0$]
    {\includegraphics[width=0.2\textwidth]{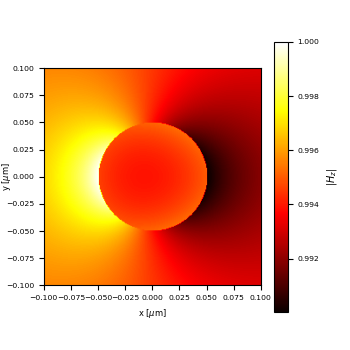}} 
    \newline
    \captionsetup[subfigure]{justification=centering}
    \subfloat[mode 4, $\varepsilon_{di} \neq 0$]
    {\includegraphics[width=0.2\textwidth]{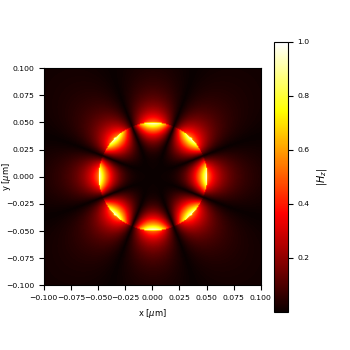}}
    \hspace{1mm}
    \captionsetup[subfigure]{justification=centering}
    \subfloat[mode 4, $\varepsilon_{di} = 0$]
    {\includegraphics[width=0.2\textwidth]{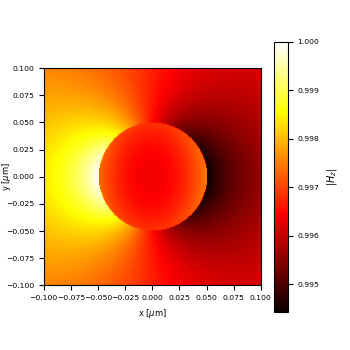}} 
    \caption{$|Hz|$ field with \fourDisp}
    \label{fig fields disp}
\end{figure}

In Fig. \ref{fig fields disp} we  give color maps of the spatial distribution of the magnetic field 
for the first four modes. The right column corresponds to the case of a core without gain ($\varepsilon_{di}=0$), whereas the left column corresponds to the case of an active core  ($\varepsilon_{di} \neq 0$). 
The values of $\varepsilon_{di} \neq 0$ have been taken to obtain almost full loss  compensation for each mode. 
We used the critical values $[\varepsilon_{di}]_c = -0.253806, -0.203944, -0.183008, -0.174493$ for the imaginary part of the permittivity of the active medium and for the frequencies we used the values $\omega_c/c = [0.798031, 0.894243, 0.977434, 1.05174] $  $\mu m^{-1}$ for the dipolar, quadrupolar, hexapolar and octupolar modes. We again observe that the inclusion of gain inside the cylinder  sharply highlights  the multipolar characteristics of the near field. 
%
%
%
%
%
%
%

\section{Conclusion}

We have  investigated the lasing and optical amplification conditions for the LSP modes on a cylindrical wire wrapped with graphene. Regarding the material of the cylindrical wire, two different cases have been considered: an infrared/THz transparent material and a nanocrystal (a metal-like material). While in the first case the active medium compensates plasmon losses only in the graphene monolayer, in the second case the active medium compensates  losses both in the graphene layer and in the nanocrystal. 

In a first stage we  have used an eigenmode approach to calculate the trajectories of the eigenfrequencies in the complex frequency plane when the optical gain parameter is varied. 
This procedure allowed us to obtain the critical values of the optical gain for which a modal eigenfrequency trajectory crosses the real axis. It is for these critical values that the lasing condition is fulfilled for that mode. In a second stage, we have used an scattering approach which allowed us to get a complementary understanding of plasmonic 
losses compensation and lasing conditions in terms of scattering observables such as scattering, extinction and absorption cross sections. To analysis the results we invoke analytical expressions obtained by us using the quasistatic approximation. Our findings show that the studied systems present a wide frequency range tunability of lasing resonant states. Moreover, the gain modal critical values exhibit a great dependence on chemical potential. Wires with smaller radius show much  smaller gain modal critical values. Of the studied modes, the dipolar one showed the largest gain modal critical values. Both results suggest that radiative losses are the key factor controlling the gain modal critical values.


We believe that these results provide a deeper understanding of the characteristics of LSP spasers based on graphene and will motivate further exploration of other spaser configurations exploiting the optical advantages of the graphene electromagnetic response in the infrared and terahertz ranges. This may find numerous applications in terahertz spectroscopy, terahertz imaging, or in sensing of biological samples for example, where tissues are typically transparent to the frequency range studied.

\section*{Acknowledgments} 
We acknowledge financial support by Consejo Nacional de Investigaciones Cient\'ificas y T\'ecnicas (CONICET); Secretar\'ia de Ciencia y Tecnolog\'ia de la Universidad Nacional de C\'ordoba (SECYT-UNC); and Agencia Nacional de Promoci\'on Cient\'ifica y Tecnol\'ogica (ANPCyT, PICT-2018-03587).

\section*{Disclosures}
The authors declare no conflicts of interest.



\medskip 
\bibliography{PrelatCilGraf01}

\end{document}